\documentclass[12pt]{article}
\usepackage{amssymb,latexsym,amsmath}

\newcommand{\bea}{\begin{eqnarray}}
\newcommand{\eea}{\end{eqnarray}}
\newcommand{\beaa}{\begin{eqnarray*}}
\newcommand{\eeaa}{\end{eqnarray*}}

\begin{document}

\title{\textbf{Reconstruction Procedure in Nonlocal Cosmological Models}}

\author{E.~Elizalde$^{1}$\footnote{E-mail: elizalde@ieec.uab.es, elizalde@math.mit.edu} ,
E.~O.~Pozdeeva$^{2}$\footnote{E-mail: pozdeeva@www-hep.sinp.msu.ru} , and S.~Yu.~Vernov$^{1,2}$\footnote{E-mail: vernov@ieec.uab.es, svernov@theory.sinp.msu.ru}  \vspace*{3mm} \\
\small $^1$Instituto de Ciencias del Espacio (ICE/CSIC) \, and \\
\small  Institut d'Estudis Espacials de Catalunya (IEEC), \\
\small  Campus UAB, Facultat de Ci\`encies, Torre C5-Parell-2a planta, \\
\small  E-08193, Bellaterra (Barcelona), Spain\\
\small  $^2$Skobeltsyn Institute of Nuclear Physics,  Lomonosov Moscow State University,\\
\small  Leninskie Gory 1, 119991, Moscow, Russia
}
\date{ }

\maketitle

\begin{abstract}
A nonlocal gravity model is considered which does not assume the existence of a new dimensional parameter in the action and includes a function $f(\Box^{-1} R)$, with $\Box$ the d'Alembertian operator.
Using a reconstruction procedure for the local scalar-tensor formulation of this model, a function $f$ is obtained for which the model exhibits power-law solutions with the Hubble parameter $H=n/t$, for any value of the constant $n$. For generic $n$---namely except for a few special values which are characterized and also specifically studied---the corresponding function $f$ is a sum of exponential functions. Corresponding power-law solutions are found explicitly. Also the case is solved in all detail of a function $f$ such that the model contains both de Sitter and power-law solutions.
\end{abstract}


\section{Introduction}

A large number of observational data~\cite{Perlmutter:1998np,Riess:1998cb,WMAP,Tegmark,Seljak:2004xh,Jain:2003tba} supports that the expansion of the Universe is accelerating now and has been doing so for a considerable part of its existence.
The assumption that General Relativity is the
correct theory of gravity leads to the conclusion that about seventy percents of the energy density of the present day Universe
should be a smoothly distributed, slowly varying cosmic fluid called dark energy~\cite{DE_rev}. If, alternatively, one wants to describe the accelerating expansion
without this fluid, the gravity equation must be modified (for recent reviews, see, e.g.,~\cite{NO-rev,Review-Nojiri-Odintsov,Book-Capozziello-Faraoni}).  There has been a lot of discussion on which side of the Einstein equations should be modified, either by adding
the cosmological constant, or anything similar,
to the matter side of the equation, or by modifying the
gravity side of the equation~\cite{Eli0408}. Note that some of these modified gravity models, for example, $F(R)$ gravity models, can be transformed into
general relativity  with additional scalar fields by a suitable conformal transformation of the metric~\cite{Book-Capozziello-Faraoni,Felice_Tsujikawa}.

Higher-derivative corrections to the Einstein--Hilbert action are being actively studied
in the context of quantum gravity (as one of the first papers we can mention~\cite{Stelle}, see also~\cite{BOSh} and references therein).
A nonlocal gravity theory
obtained by taking into account quantum effects has been proposed
in~\cite{Deser:2007jk}. In a way, this development goes back to string/M-theory, which is
usually considered as a possible theory to encompass all known fundamental
interactions, including gravity. The appearance of nonlocality
within string field theory provides therefore a good motivation for studying nonlocal
cosmological models.

The majority of available nonlocal cosmological models
explicitly include a function of the d'Alem\-bertian operator, $\Box$, and
either define a nonlocal modified
gravity~\cite{Deser:2007jk,Non-local-gravity-Refs,Odintsov0708,Jhingan:2008ym,Koivisto:2008,Koivisto2,Woodard,Non-local-FR,
Nojiri:2010pw,Bamba1104,ZS,EPV2011,EPV_PoS,Vernov1202} or contain a nonlocal scalar field,
minimally coupled to gravity~\cite{Non-local_scalar}.

In this paper, we consider a nonlocal gravity model that includes a function of
the $\Box^{-1}$ operator and does not assume the existence of a new dimensional parameter in the action~\cite{Deser:2007jk}. The considered nonlocal model has a local scalar-tensor formulation~\cite{Odintsov0708}. As was shown
in~\cite{Odintsov0708}, a theory of this kind may actually encompass all known sequence of the Universe
history: inflation, radiation/matter dominance and a subsequent and final dark
epoch. In~\cite{Koivisto:2008} the ensuing cosmologies corresponding to the four basic epochs:
radiation dominance, matter dominance, acceleration, and a general
scaling law have been studied for nonlocal models involving, in particular, an
exponential form for the function $f(\psi)$. The perturbation analysis of this model has been developed and
the Solar System test has been done in~\cite{Koivisto2}.
 An explicit mechanism to screen the cosmological constant in
nonlocal gravity was discussed in~\cite{Nojiri:2010pw,Bamba1104,ZS}.

 The cosmological model often includes some function which
cannot be deduced from the fundamental theory, usually one can only assume some property of this function, for example, that the function should be a polynomial or more generally an analytic function. Then the question arises, why some specific form of the function $f(\Box^{-1})$ is chosen and not a different one, what is the physical motivation for the final choice. To answer these questions one has to resort to the reconstruction procedure itself and to understand in depth the expansion history of the Universe in order to match both with care. For this nonlocal model, a technique for choosing the distortion function $f(\Box^{-1})$,
so as to fit an arbitrary expansion history has been derived in~\cite{Woodard}.
In the framework of the local formulation, a reconstruction procedure
has been proposed~\cite{Koivisto:2008} in terms of functions of the scale
factor~$a$. The  reconstruction method considered in this paper was first proposed in~\cite{EPV2011} with the aim to obtain models with de Sitter solutions and it has been generalized in~\cite{Vernov1202}. This method uses convenient functions of the cosmic time $t$.

In the present paper, we develop in some detail the reconstruction procedure and use it to find models with power-law solutions.
We obtain functions $f$ which correspond to power-law solutions, e.g., solutions for which the Hubble parameter is given by $H=n/t$, where $n$ is a real number. Also, we find functions $f$ for which the model has both the de Sitter and power-law solutions.
It should be noted that the reconstruction procedure is widely used in cosmology, in particular,
in models with minimally coupled scalar fields~\cite{SRSS,scalarfields,Superpotential,NoOdin2007}, also non-minimally coupled scalar~\cite{KTV11042125} or Yang--Mills~\cite{BNO} fields, $F(R)$ and Gauss--Bonnet gravity models~\cite{NoOdin2007,FR_reconsruction}, and  $F(T)$ models, with $T$ being the torsion scalar~\cite{FT_Reconstruction}.

The paper is organized as follows.
In Sect.~2, we consider the action for a general class of nonlocal gravity
models and derive the corresponding equations of motion.
In Sect.~3 we propose the algorithm which allows to find $f(\psi)$ and get the model with the given cosmological evolution. In Sect.~4, we prove that the model with $f(\psi)$ being a sum of
exponential functions can indeed lead to power-law solutions. This is extended in Sect.~5
where we find sufficient conditions on $f(\psi)$, under which the model has power-law solutions, and present these solutions in the explicit form.
In Sect.~6 a function $f(\psi)$ is considered which corresponds to a model with both de Sitter and power-law solutions, what is a quite interesting, somehow unexpected finding.
In Sect.~7, we consider a few special cases of power-law solutions and get for each case the corresponding function $f(\psi)$.
In Sect.~8, we consider the reconstruction procedure for the model with many perfect fluids.
The last section is devoted to conclusion.

\section{Scalar-tensor gravity model}

We consider a class of nonlocal gravities, whose action is given by
\begin{equation}
\label{action}
S=\int d^4 x \sqrt{-g}\left\{
\frac{1}{2\kappa^2}\left[ R\left(1 + f(\Box^{-1}R)\right)
-2\Lambda \right] + \mathcal{L}_\mathrm{matter}
\right\} ,
\end{equation}
where $\kappa^2=8\pi G=8\pi/M_{\mathrm{Pl}}^2$, the Planck mass
being $M_{\mathrm{Pl}} = G^{-1/2} = 1.2 \times 10^{19}$ GeV, while $f$ is a differentiable  function that characterizes the
nature of nonlocality, with $\Box^{-1}$ being the inverse of
the d'Alembertian operator, $\Lambda$ is the cosmological constant, and $\mathcal{L}_\mathrm{matter}$
the matter Lagrangian.   We use the
signature $(-,+,+,+)$, $g$ being the determinant of the metric tensor
$g_{\mu\nu}$. Recall the covariant d'Alembertian for a scalar field, which reads
\begin{equation*}
\Box\equiv \nabla^\mu\nabla_\mu =\nabla^\mu\partial_\mu= \frac{1}{\sqrt{-g}} \partial_\mu \left(
\sqrt{-g} \, g^{\mu
  \nu}\partial_\nu \right),
\end{equation*}
where $\nabla^\mu$ is the covariant derivative.

Introducing two scalar fields, $\psi$ and $\xi$, one can rewrite
action~(\ref{action}) as follows
\begin{equation}
\label{action2} S= \int d^4x \sqrt{-g}\left\{
\frac{1}{2\kappa^2}\left[R\left(1 + f(\psi)+\xi\right) -  \xi \Box
\psi - 2 \Lambda \right]+ \mathcal{L}_\mathrm{matter} \right\}\!.
\end{equation}
By varying action (\ref{action2}) over $\xi$, we get
$\Box\psi=R$.
Substituting $\psi=\Box^{-1}R$ into action~(\ref{action2}), one reobtains
action~(\ref{action}). Action (\ref{action2})
may thus be regarded as a local action.

By varying this action with respect to $\xi$ and
$\psi$, respectively, one obtains the field equations
\bea
\label{equpsi}
\Box\psi&=&R\,,
\\
\Box\xi&=& f'(\psi) R\, ,
\label{equxi}
\eea
 where the prime denotes derivative with respect to $\psi$.

Varying action~(\ref{action2}) with respect to the metric tensor
$g_{\mu\nu}$ yields
\begin{equation}
\label{nl4}
\frac{1}{2}g_{\mu\nu} \left[R\Psi
 + \partial_\rho \xi \partial^\rho \psi - 2 (\Lambda+\Box\Psi) \right]
 - R_{\mu\nu}\Psi-\frac{1}{2}\left(\partial_\mu \xi \partial_\nu \psi
+ \partial_\mu \psi \partial_\nu \xi \right)
  + \nabla_\mu \partial_\nu\Psi=- \kappa^2T_{\mathrm{m}\, \mu\nu}\, ,
\end{equation}
where $\Psi=1+f(\psi) +\xi$, the energy--momentum tensor of matter $T_{\mathrm{m}\,\mu\nu}$ is
defined as
\begin{equation}\label{Tmatter}
T_{\mathrm{m}\, \mu\nu} \equiv
-\frac{2}{\sqrt{-g}}
 \frac{\delta \left(\sqrt{-g} {\mathcal{L}}_{\mathrm{matter}}\right)}{\delta g^{\mu \nu}}.
\end{equation}

The system of equations (\ref{nl4})
does not include the function $\psi$, but only $f(\psi)$ and $f'(\psi)$, together
with derivatives of $\psi$. Note that $f(\psi)$ can be determined up to a constant, because one can add a constant to $f(\psi)$ and subtract the same constant from $\xi$, without changing any
of the equations.

To consider cosmological solutions we assume a spatially
flat Friedmann--Lema\^{i}tre--Ro\-bertson--Walker (FLRW) universe, with the interval,
\begin{equation}
ds^2=-dt^2+a^2(t)\delta_{ij}dx^idx^j\,,
\end{equation}
and consider scalar fields which depend on time only.
In the FLRW metric, the system of Eqs.~(\ref{equpsi})--(\ref{nl4})
reduces to
\bea \label{einstein1}
 3H^2\Psi&=& \!{}- \frac{1}{2}\dot\xi \dot\psi
 - 3H\dot\Psi  + \Lambda
+ \kappa^2 \rho_{\mathrm{m}}\, ,\\
\label{einstein2} \left(2\dot H + 3H^2\right)\Psi &=&
\frac{1}{2}\dot\xi \dot\psi -
\ddot\Psi- 2H \dot\Psi + \Lambda - \kappa^2 P_{\mathrm{m}}\,,\\
\label{xieq} \ddot \xi  &=& \!{}- 3H \dot \xi - 6\left( \dot H + 2
H^2\right)f'(\psi) \,,\\
\label{psieq}\ddot \psi &=&  \!{}- 3H \dot \psi - 6 \left( \dot H + 2
H^2\right),
\eea
where  differentiation with
respect to time $t$ is denoted by a dot, and the Hubble parameter is
$H=\dot a/a$. We assume that matter is a perfect fluid with
$T_{\mathrm{m}\, 0 0} = \rho_{\mathrm{m}}$ and
$T_{\mathrm{m}\, i j} = P_{\mathrm{m}} g_{i j}$. The continuity equation reads
\begin{equation}
\label{equ_rho} \dot\rho_{\mathrm{m}}={}-
3H(P_{\mathrm{m}}+\rho_{\mathrm{m}}).
\end{equation}
Adding up (\ref{einstein1}) and (\ref{einstein2}), we obtain
the following second order linear differential equation for~$\Psi$:
\begin{equation}
\label{equPsi} \ddot\Psi+5H\dot\Psi+\left(2\dot H +
6H^2\right)\Psi- 2\Lambda +\kappa^2
(P_{\mathrm{m}}-\rho_{\mathrm{m}})=0.
\end{equation}

In the case when $\Lambda=0$ and matter is absent, Eq.~(\ref{equPsi}) has a solution $\Psi(t)\equiv 0$.
As one can see from Eq.~(\ref{einstein1}), in this case functions $\psi$ and $\xi$ are both constants.
From Eq.~(\ref{equpsi}), it follows that the space-time is Ricci flat. Therefore, the Minkowski space is a solution in this case.

\section{Nonlocal gravitational models with a given cosmological evolution}

Let us now consider the question, how to get the function $f(\psi)$ that
corresponds to a solution with a given Hubble parameter $H(t)$?
If the equation
of state (EoS) parameter $w_{\mathrm{m}}$ is known, then for any given $H(t)$ one can integrate Eq.~(\ref{equ_rho}) and get $\rho_{\mathrm{m}}(t)$, after which one can solve
Eq.~(\ref{equPsi}) and obtain $\Psi(t)$. The knowledge of $H(t)$ also allows to integrate (\ref{equpsi}) and get $\psi(t)$.

Substituting
\begin{equation*}
\xi(t)=\Psi(t)-f(\psi)-1
\end{equation*}
 into Eq.~(\ref{xieq}) and using the function $t(\psi)$,
we get a linear differential equation for $f(\psi)$, namely
\begin{equation}
\label{equaf} \dot\psi^2 f''(\psi)-12\left(\dot H + 2 H^2\right)f'(\psi)=\ddot\Psi+3H\dot\Psi\,.
\end{equation}
 It can be more suitable for the calculation to rewrite Eq.~(\ref{equaf}) as follows:
\begin{equation}
\label{s}
\dot s\dot\psi-12\left(\dot H+2H^2\right)s=\ddot\Psi+3H\dot\Psi\,,
\end{equation}
where $s(t)=f'(\psi(t))$. From this first order differential equation one can obtain the function $s(t)$, and, after that, get $f'(\psi)$ and $f(\psi)$.

Note that Eq.~(\ref{equaf}) is a necessary condition for the model to have the
solutions in the given form. Substituting the obtained function $f(\psi)$ into Eq.~(\ref{einstein1}) or Eq.~(\ref{einstein2}), one can check the existence of solutions of the given
form. Note that this method not only allows to construct the corresponding $f(\psi)$, but it also gives a suitable way to obtain $\xi(t)$ and $\psi(t)$ for a given $H(t)$. In the following Sections we will illustrate this result for some physically important examples of possible $H(t)$ behaviors.

\section{Power-law solutions}

In~\cite{EPV2011} we have found the general form of the function $f(\psi)$ for which the model can have de Sitter solutions. Here, we explore power-law solutions, in other words, we assume that the Hubble parameter is of the form $H=n/t$.

In the General Relativity, a power-law solution $H=n/t$ corresponds to model with perfect fluid,
 and the EoS
parameter $w_{\mathrm{m}}\equiv P_{\mathrm{m}}/\rho_{\mathrm{m}}$ is equal to $w_{\mathrm{m}}={}-1+2/(3n)$ at that.
In the considering model, we assume that the matter has the EoS
parameter $w_{\mathrm{m}}$ being an arbitrary
constant, not equal to $-1$.

For  $H=n/t$, Eq.~(\ref{equ_rho}) has the following general solution:
\begin{equation}
\label{rho_sol}
\rho_{\mathrm{m}}(t)=\rho_0t^{-3n(w_{\mathrm{m}}+1)},
\end{equation}
where $\rho_0$ is an arbitrary constant. Equation~(\ref{equPsi}) has the following general solution:
\begin{itemize}
\item
For $n\neq -1$ and $n \neq -1/3$
\begin{equation}
\label{solPsi1}
   \Psi_0=C_1t^{-2n}+C_2t^{1-3n}+\frac{\Lambda}{(n+1)(3n+1)} t^2-\frac{\rho_0 \kappa^2(w_{\mathrm{m}}-1)t^{2-3(1+w_{\mathrm{m}})n}}{(3nw_{\mathrm{m}}-1)(n+3nw_{\mathrm{m}}-2)}.
\end{equation}
\item
For $n=-1$,
\begin{equation}
\label{solPsi2}
  \Psi_1=C_1t^{2}+C_2t^{4}-\Lambda t^2\left(\ln(t)+\frac{1}{2}\right)-\frac{\rho_0 \kappa^2(w_{\mathrm{m}}-1)}{3(w_{\mathrm{m}}+1)(3w_{\mathrm{m}}+1)}t^{5+3w_{\mathrm{m}}}.
\end{equation}
\item
For $n=-1/3$,
\begin{equation}
\label{solPsi3}
  \Psi_2=C_1t^{2}+C_2t^{2/3}+\frac{3}{2}\Lambda t^2\left(\ln(t)-\frac{3}{4}\right)
-\frac{3\rho_0 \kappa^2(w_{\mathrm{m}}-1)}{(w_{\mathrm{m}}+1)(3w_{\mathrm{m}}+7)}t^{3+w_{\mathrm{m}}}\,,
\end{equation}
\end{itemize}
where $C_1$ and $C_2$ are arbitrary constants. The EoS parameter
 $w_{\mathrm{m}}$ is chosen so that
$(3nw_{\mathrm{m}}-1)(n+3nw_{\mathrm{m}}-2)$ be not equal to zero.

Inserting $H=n/t$ into  Eq.~(\ref{psieq}) and assuming $n\neq 1/3$ and $n\neq 1/2$,
we find the  solution for $\psi(t)$ to be
\begin{equation}
\label{psisoln}
\psi(t)=\psi_1t^{1-3n}-\frac{6n(2n-1)}{3n-1}\ln{\left(\frac{t}{t_0}\right)}\,,
\end{equation}
where $\psi_1$ and $t_0$ are arbitrary  constants.  In all formulae we assume that $t>0$, so, $t_0>0$.
Also, we get that the general solution of (\ref{psieq}) is
\begin{eqnarray}
\psi(t)&=&C_3
t^{-1/2}+C_4,\quad \mbox{at} \quad n=\frac{1}{2},\label{1/2psi}\\
\psi(t)&=&\frac{1}{3}\left(\ln(t)\right)^2+\tilde{C}_3\ln(t)+\tilde{C}_4,\quad
\mbox{at}\quad n=\frac{1}{3},\label{1/3psi}
\end{eqnarray}
where $C_3$, $C_4$, $\tilde{C}_3$, and $\tilde{C}_4$ are arbitrary constants.
We assume that $\psi(t)$ is not a constant, so $C_3\neq 0$.

To get $f(\psi)$ in terms of elementary functions for any $n$, except for the above-mentioned particular values,  it is useful
to set $\psi_1=0$ in (\ref{psisoln}) and to express the time variable $t$ in terms of $\psi$, as\footnote{If $\psi_1\neq0$, then $t(\psi)$ includes the Lambert function,  so in this case, we do not get $f(\psi)$ in terms of elementary functions.}
\begin{equation}
\label{tpsi}
t=t_0 e^{(1-3n)\psi/(6n(2n-1))}.
\end{equation}
We obtain the following general solution to (\ref{equaf}):
\begin{equation}
\label{fsumexp}
\begin{split}
 f_0(\psi)&=\frac{\Lambda t_0^2}{6n(1+n)}e^{(1-3n)\psi/(3n(2n-1))}-\frac{\rho_0\kappa^2t_0^{2-3n-3nw_{\mathrm{m}}}}{3n(n-2+3nw_{\mathrm{m}})}
e^{(3n(1+w_{\mathrm{m}})-2)(3n-1)\psi/(6n(2n-1))}
\\&{}+\frac{(n-1)t_0^{-2n}C_1}{2(2n-1)}e^{(3n-1)\psi/(3(2n-1))}
+D_1e^{(3n-1)^2\psi/(3n(2n-1))}+D_2,
\end{split}
\end{equation}
where $D_1$ and $D_2$ are integration constants. Note that $C_1$ is an arbitrary constant as well.
To get this result we assume that $n\neq 1/2$,  $n\neq 1/3$, $n\neq -1/3$, and $n\neq -1$.

Let us consider the case $n=-1$. Substituting $\Psi(t)$, given by (\ref{solPsi2}), into Eq.~(\ref{equaf}), we get the following general solution:
\begin{equation}
\label{psiexppsi}
\begin{split}
f_1(\psi)&=\frac{C_1}{3}t_0^2 e^{4\psi/9} -{\frac {\Lambda}{54}}
\left( 18\ln(t_0)+3+4\psi \right) t_0^2{e^{4\psi/9}}\\&{}-\frac{\rho_0\kappa^2t_0^{9(5+3w_{\mathrm{m}})}}{w_{\mathrm{m}}+1}e^{2(
5+3w_{\mathrm{m}})\psi/9}+D_1e^{16\psi/9}+D_2.
\end{split}
\end{equation}
Note that this form of $f(\psi)$ is similar to a function for which there exist de Sitter solutions at $w_{\mathrm{m}}=1/3$ (see Sect.~6). At the same time, $f_1(\psi)$ is a sum of exponential functions provided $\Lambda=0$.

For $n=-1/3$, we use (\ref{solPsi2}) and get
\begin{equation*}
f_2(\psi)=\frac25\,C_1 t_0^{2/3}{e^{2\psi/5}}-\frac34\,t_0^2
\Lambda e^{6\psi/5}-3\frac
{\rho_0 \kappa^2t_0^{3+w_{\mathrm{m}}}}{3w_{\mathrm{m}}+7}e^{3\left(
3+w_{\mathrm{m}} \right)\psi/5}+D_1e^{12\psi/5}+D_2.
\end{equation*}
Note that $f_2(\psi)$ coincides with $f_0(\psi)$ for $n=-1/3$, so
formula (\ref{fsumexp}) can be used for $n=-1/3$ as well.

The cases $n=1/3$ and $n=1/2$, where the function $\psi(t)$ has a special form, different from~(\ref{psisoln}), will be considered in Sect.~7.

We see that for all values of $n$, except when $n=-1$, $n=1/3$, and $n=1/2$,  $f(\psi)$ is equal to either an exponential function or to a sum of exponential functions plus a constant.
This is the general form of the function $f(\psi)$, which can have power-law solutions in the model with
the cosmological constant and an ideal perfect fluid. We have not proven that this model has power-law solutions, what we really proved is that other models do not have such solutions (with $\psi_1=0$).

\section{Power-law solutions for a given $f(\psi)$}

The full system of equations is (\ref{einstein1})--(\ref{equ_rho}). We solve equations (\ref{xieq})--(\ref{equ_rho}) and
the sum of equations~(\ref{einstein1}) and (\ref{einstein2}). Thus, to get the solution it is enough to substitute the functions obtained into Eq.~(\ref{einstein1}).

After this substitution, we get that terms which are proportional to $\Lambda$ and $\rho_0$ cancel and that Eq.~(\ref{einstein1}) takes the following form:
\begin{equation}
\label{cond}
6t_0^{2(3n-1)}(2n-1)nD_1t^{-6n}=0, \quad\Rightarrow\quad  D_1=0.
\end{equation}
Note that the case $n=1/2$ has been excluded (in this case the function $\psi(t)$ has another form).
Thus, we get that the model with
\begin{equation}
\label{fpsi}
f(\psi)=\tilde{f}_1e^{\alpha_1^{\phantom{1}}\psi}+\tilde{f}_2e^{\alpha_2^{\phantom{1}}\psi}+
\tilde{f}_3e^{\alpha_3^{\phantom{1}}\psi}+\tilde{f}_4,
\end{equation}
where $\tilde{f}_i$ and $\alpha_i$ are constants, has solutions with $H=n/t$, where $n$ is an arbitrary real number, but $n\neq -1$, $n\neq 1/3$ and $n\neq 1/2$. As we have noted in Sect.~2, the function $f(\psi)$ is defined up to a constant, so we can put $\tilde{f}_4=0$ without loss of generality.

The method here considered allows not only to get the suitable function $f(\psi)$, but also to obtain solutions in explicit form. Indeed, we have found that the  model considered, with $f(\psi)$ given by (\ref{fpsi}), has the  solution
\begin{equation}\label{solutions}
\begin{split}
H(t)&=\frac{n}{t},\\
\rho_{\mathrm{m}}(t)&=\rho_0t^{-3n(w_{\mathrm{m}}+1)},\\
\psi(t)&={}-\frac{6n(2n-1)}{3n-1}\ln{\left(\frac{t}{t_0}\right)},\\
\Psi(t)&=\left\{\!\!\!
\begin{array}{l}
\displaystyle \Theta+\frac{\Lambda t^2}{(n+1)(3n+1)}, \quad n\neq{}-\frac13\,,\\
\displaystyle \Theta+\frac{3}{2}\Lambda t^2\left(\ln(t)-\frac{3}{4}\right), \quad n={}-\frac13\,,
\end{array}\right.\\
\xi(t)&=\Psi(t)-f(\psi)-1,
\end{split}
\end{equation}
where
\begin{equation*}
\Theta\equiv C_1t^{-2n}+C_2t^{1-3n}-\frac{\rho_0 \kappa^2(w_{\mathrm{m}}-1)t^{2-3(1+w_{\mathrm{m}})n}}{(3nw_{\mathrm{m}}-1)(n+3nw_{\mathrm{m}}-2)}\,,
\end{equation*}
the constants being subject to the following conditions:
\begin{equation}
\label{fcoefs}
\begin{array}{ll}
\displaystyle \tilde{f}_1=\frac{\Lambda t_0^2}{6n(1+n)},&\displaystyle \alpha_1=\frac{1-3n}{3n(2n-1)},\\[2.7mm]
\displaystyle \tilde{f}_2={}-\frac{\rho_0\kappa^2t_0^{2-3n-3nw_{\mathrm{m}}}}{3n(n-2+3nw_{\mathrm{m}})},
&\displaystyle \alpha_2=\frac{(3n(1+w_{\mathrm{m}})-2)(3n-1)}{6n(2n-1)},\\[2.7mm]
\displaystyle \tilde{f}_3=\frac{(n-1)t_0^{-2n}C_1}{2(2n-1)}, &\displaystyle \alpha_3=\frac{3n-1}{3(2n-1)}.
\end{array}
\end{equation}
Thus, for the model with the given $f(\psi)$ and $\Lambda$, the values of the parameters
$C_1$ and $t_0$ are fixed and there is a connection between $\alpha_1$ and $\alpha_3$.
The values of  $\alpha_2$ and $\tilde{f}_2$  define conditions on  $w_{\mathrm{m}}$, which is not equal to $-1$, and $\rho_0$.
Note that the solutions  include an arbitrary parameter~$C_2$.
Recall also that $n\neq -1$, $n\neq 1/3$ and $n\neq 1/2$, and we demand that
$n-2+3nw_{\mathrm{m}}\neq 0$ and $w_{\mathrm{m}}\neq -1$. As a consequence of these results, it follows that the model without matter can develop power-law solutions only if $\tilde{f}_2=0$, thus $f(\psi)$ is either an exponential function or a sum of two exponentials.

\section{Models with both power-law and de Sitter solutions}
\subsection{The function $f(\psi)$ for the model with de Sitter solutions}
 In~\cite{EPV2011} we obtained that the model admits the de Sitter solution $H=H_0$, where $H_0$ is a nonzero constant, provided that
\begin{equation}
\label{fdeSitter}
   f_{dS}(\psi)
=c_1e^{\psi/2}+c_2e^{3\psi/2}-\frac{\kappa^2\tilde{\rho}_0}{3(1+3w_{\mathrm{m}})H_0^2}
e^{3(w_{\mathrm{m}}+1)\psi/4}\,, \quad \mbox{for}\quad w_{\mathrm{m}}\neq {}-\frac{1}{3}\,,
\end{equation}
\begin{equation}
\label{fdeSitter13}
   \tilde{f}_{dS}(\psi)
=c_1e^{\psi/2}+c_2e^{3\psi/2}+\frac{\kappa^2\tilde{\rho}_0}{4H_0^2}\left(1-\frac{1}{3}\psi\right)
e^{\psi/2}, \quad \mbox{for}\quad w_{\mathrm{m}}= {}-\frac{1}{3}\,,
\end{equation}
where $c_1$  and $c_2$ are arbitrary constants, and $\tilde{\rho}_0$ is the initial value of the matter energy density.
Note that we include in this model only matter with the same EoS parameter $w_{\mathrm{m}}$, but that the initial values of the matter energy density, $\rho_0$ and $\tilde{\rho}_0$, can be actually different.

In~\cite{EPV2011}, a de Sitter solution in its most general form has been found for $f(\psi)$ being an exponential function. In particular, one gets that equations (\ref{psieq})--(\ref{equPsi}), which do not depend on the function $f(\psi)$, have the following  solutions:
\begin{equation}
\label{dSsol}
 \begin{split}
H(t)&=H_0,\\
\rho_{\mathrm{m}}(t)&= \tilde{\rho}_0\,e^{{}-3(1+w_{\mathrm{m}})H_0(t-\tilde{t}_0)},\\
\psi(t)&= {}-4H_0(t-\tilde{t}_0),\\
\displaystyle \Psi(t)&= \left\{\!\!\!
\begin{array}{l}
\displaystyle \Upsilon-\frac{\kappa^2\tilde{\rho}_0(w_{\mathrm{m}}-1)}
{3H_0^2w_{\mathrm{m}}(1+3w_{\mathrm{m}})}e^{-3H_0(w_{\mathrm{m}}+1)(t-\tilde{t}_0)}\,,\qquad w_{\mathrm{m}}\neq 0,\quad  w_{\mathrm{m}}\neq \frac{1}{3},\\[2.7mm]
\displaystyle \Upsilon
-\frac{\kappa^2\tilde{\rho}_0}{H_0}e^{-3H_0(t-\tilde{t}_0)}(t-\tilde{t}_0),
\qquad w_{\mathrm{m}}=0,\\[2.7mm]
\displaystyle \Upsilon+\frac{4\kappa^2\tilde{\rho}_0}{3H_0}e^{-2H_0(t-\tilde{t}_0)}(t-\tilde{t}_0),\qquad w_{\mathrm{m}}=-\frac{1}{3},
\end{array}\right.
\end{split}
\end{equation}
where
\begin{equation}
\Upsilon=c_3e^{-3H_0(t-\tilde{t}_0)}+4c_1e^{-2H_0(t-\tilde{t}_0)}+\frac{\Lambda}{3H_0^2},
\end{equation}
$c_3$ and $\tilde{t}_0$ being arbitrary constants.

Substituting into Eq.~(\ref{einstein1}) solution (\ref{dSsol}) and $\xi(t)=\Psi(t)-f(\psi)-1$, where $f(\psi)$ is specified by (\ref{fdeSitter}) or (\ref{fdeSitter13}), we get
\begin{equation*}
12c_2H_0^2e^{-6H_0(t-\tilde{t}_0)}=0, \quad\Rightarrow\quad c_2=0.
\end{equation*}
Note that we obtain the same results for all values of $w_{\mathrm{m}}$.
Thus, the model has de Sitter solutions if
\begin{equation}
\label{fdeSitterpowerlaw}
 f_{dS}(\psi)
=c_1e^{\psi/2}-\frac{\kappa^2\tilde{\rho}_0}{3(1+3w_{\mathrm{m}})H_0^2}
e^{3(w_{\mathrm{m}}+1)\psi/4}\,, \quad \mbox{for } \  w_{\mathrm{m}}\neq {}-\frac{1}{3}\,,
\end{equation}
\begin{equation}
\label{fdeSitter13s} \tilde{f}_{dS}(\psi)
=c_1e^{\psi/2}+\frac{\kappa^2\tilde{\rho}_0}{4H_0^2}\left(1-\frac{1}{3}\psi\right)
e^{\psi/2}, \quad \mbox{for } \  w_{\mathrm{m}}= {}-\frac{1}{3}\,,
\end{equation}
with arbitrary $c_1$.

Let us check for the possibility that the model, with some given function $f(\psi)$, has both de Sitter and power-law solutions.
Such possibility exists only in the case $w_{\mathrm{m}}\neq {}-1/3$.

Indeed, from (\ref{fdeSitter13s}) it is follow that for $w_{\mathrm{m}}=-1/3$ de Sitter solutions exist for the function
\begin{equation}
\label{fpsi3}
f(\psi)=\tilde{f}_1e^{\beta_1\psi}+\tilde{f}_2e^{\beta_2\psi}\psi,
\end{equation}
with $\beta_2=1/2$ and $\tilde{f}_2\neq 0$. A power-law solution does not exist for such a function $f(\psi)$.
It is obvious from (\ref{psiexppsi}), that the function $f(\psi)$ with the term $\tilde{f}_2e^{\beta_2\psi}\psi$
can has a power-law solution only at $\beta_2=4/9$.
Therefore, a model of this type cannot have both de Sitter and power-law solutions.

We thus  conclude that the function $f(\psi)$ should be either a single exponential function or a sum of two exponential functions. Let us now check these two possibilities separately.

\subsection{A single exponential $f(\psi)$}
Let us consider the model with
\begin{equation}\label{fexpo}
f(\psi)=\tilde{f}_0e^{\beta \psi},
\end{equation}
where $\tilde{f}_0$ an $\beta$ are constants.
This case where $f(\psi)$ is an exponential function has been mostly investigated~\cite{Odintsov0708,Jhingan:2008ym,Koivisto:2008,Nojiri:2010pw,Bamba1104,ZS,EPV2011,EPV_PoS,EPVZ2012}.
Particular de Sitter and power-law solutions were found in~\cite{Odintsov0708} and, after that,
de Sitter solutions have been studied in~\cite{Bamba1104,EPV2011,EPV_PoS} and power-law solutions in~\cite{ZS,EPVZ2012}, but always separately. Using mainly results in~\cite{EPV2011,EPVZ2012}, our aim here will be to construct a model which simultaneously exhibits both kinds of solutions: de Sitter and power-law.

If we assume that the matter is absent, then it follows from (\ref{fdeSitterpowerlaw}) that the model has de Sitter solutions only in the case when
\begin{equation*}
\beta=\frac{1}{2}.
\end{equation*}
It is easy to see that $\alpha_3\neq 1/2$, for any $n$, so that the model can have power-law solutions only if $\alpha_1=1/2$.
In this case,
\begin{equation}
\label{n1}
n = {}-\frac{1}{4}\pm \frac{\sqrt{57}}{12}.
\end{equation}
Thus, we come to the conclusion that for $\rho(t)=0$ the model has both de Sitter and power-law solutions if and only if
$\beta=1/2$ and $\Lambda\neq 0$. Note that $\tilde{f}_0$ can be arbitrary and does not get fixed by the value of $\Lambda$.

Let us consider the opposite case, when matter is present. From (\ref{fdeSitterpowerlaw}) we get the condition of the existence of de Sitter solutions:
\begin{equation}
\beta=\frac{3}{4}(w_{\mathrm{m}}+1).
\end{equation}

If we also require that this model should have power-law solutions as well, then we get from~(\ref{fcoefs})
and condition $\alpha_2=\beta$ that
\begin{equation}
\label{n}
n = \frac{4}{3(3-w_{\mathrm{m}})}.
\end{equation}

Note that the equation $\alpha_1=\alpha_2$ with the $n$, given by (\ref{n}), has the unique solution $w_{\mathrm{m}}=-1$.
It really means that there is no matter, different from the cosmological constant.
As a consequence, these models with matter can admit power-law solutions only if $\Lambda=0$.

It has been found in~\cite{EPVZ2012} that, in the case $\Lambda=0$, power-law solutions
exist only if either $n=1/(3w_{\mathrm{m}})$ or if $n=1$ and $w_{\mathrm{m}}=-1/3$. Comparing these condition with (\ref{n}), we obtain that the model has both de Sitter and power-law solutions if and only if $n=1/(3w_{\mathrm{m}})$ and $w_{\mathrm{m}}=3/5$. In this case, we get
$\beta=6/5$ and $n = 5/9$.

Thus, we have reached the conclusion that the model with a single exponential function has both de Sitter and power-law solutions provided that either $\rho(t)=0$ or $\Lambda=0$ (but not both at once). In the first case, $\beta=1/2$. In the second one, $\beta=6/5$ and $w_{\mathrm{m}}=3/5$.

\subsection{A sum of two exponential functions}

Let us consider de Sitter and power-law solutions for
\begin{equation}
\label{fpsi2}
f(\psi)=\tilde{f}_1e^{\beta_1\psi}+\tilde{f}_2e^{\beta_2\psi}.
\end{equation}
We have shown that de Sitter solutions for such $f(\psi)$ exist only when matter is present.
From~(\ref{fdeSitterpowerlaw}), we obtain
\begin{equation}
\label{fcoefs2}
\beta_1=\frac{1}{2}, \qquad \beta_2=\frac{3}{4}(w_{\mathrm{m}}+1).
\end{equation}
Note that $w_{\mathrm{m}}\neq {}-1/3$, hence  $\beta_1\neq \beta_2$.

The de Sitter solution obtained is given by (\ref{dSsol}), where
\begin{equation*}
c_1=\tilde{f}_1, \qquad \tilde{\rho}_0={}-\frac{3(1+3w_{\mathrm{m}})H_0^2}{\kappa^2}\tilde{f}_2,\qquad w_{\mathrm{m}}=\frac{4}{3}\beta_2-1,
\end{equation*}
$t_0$ and $c_3$ being arbitrary parameters and $H_0$ an arbitrary nonzero constant.

Let us consider possible power-law solutions in this model. Because $\alpha_3\neq 1/2$, for any $n$, we need set $\alpha_1= 1/2$, thus the value of $n$ is fixed by (\ref{n1}). From $\alpha_2=\beta_2$ we then connect  $n$ and $w_{\mathrm{m}}$ by (\ref{n}), to get
\begin{equation}
\label{w2}
w_{\mathrm{m}}=\frac{25\pm 3\sqrt{57}}{3\pm\sqrt{57}}.
\end{equation}
The sign "$-$" corresponds to $w_{\mathrm{m}}\simeq -0.517$, what looks realistic for cosmological models, whereas the sign "$+$" corresponds to $w_{\mathrm{m}}\simeq 4.517$.

In summary, we have found that the model with $f(\psi)$ given by (\ref{fpsi2}) has both de Sitter and power-law solutions (a new and quite interesting case) if and only if
both $\Lambda$ and $\rho(t)$ are not equal to zero and $w_{\mathrm{m}}$ is given by (\ref{w2}).

\section{Special cases of power-law solutions}

In the previous sections, when looking for power-law solutions, we have ignored some special values of $n$. For completion of our search, we will consider those values of $n$ now. Being precise,
formula (\ref{fpsi}) is not valid for some (we call them here special) values of
$n$. We discuss in what follows such special values and look for the corresponding functions  $f(\psi)$.

We first consider solutions with $H=1/(2t)$, that corresponds to $R=0$. In this case, the function $\psi(t)$ is given by (\ref{1/2psi}).
Substituting this $\psi(t)$ and the function $\Psi$
in the form (\ref{solPsi1}), into~(\ref{s}), we obtain
\begin{equation}
\label{stn12}
s(t)=2\frac{C_1}{C_3\sqrt{t}}-\frac{4(3w_{\mathrm{m}}-1)}{3(3w_{\mathrm{m}}-2)C_3}\rho_0\kappa^2t^{1-3w_{\mathrm{m}}/2}-\frac{16}{15C_3}\Lambda t^{5/2}+D_1,
\end{equation}
where $D_1$ is an arbitrary constant.
Relation (\ref{1/2psi}) can be rewritten as
\begin{equation*}
\sqrt{t}=\frac{C_3}{\psi-C_4}.
\end{equation*}

Using this relation, we integrate (\ref{stn12}) and obtain
\begin{equation*}
   f(\psi)=\frac{C_1}{C_3^2}\psi^2-\left[2\frac{C_4C_1}{C_3^2}-D_1\right]\psi+D_2
+\frac {4
\Lambda C_3^4}{ 15[\psi-C_4]^{4}}+\frac{4\kappa^{2}\rho_0}{3[3w_{\mathrm{m}}-2]}\left[\frac{C_3}{\psi-C_4} \right]^{1-3w_{\mathrm{m}}}\!\!\!.
\end{equation*}
The function $f(\psi)$ is defined up to a constant, so one can set the integration constant $D_2=0$.

By direct substitution into (\ref{einstein1}), the following condition on the constants results:
\begin{equation}
C_2=C_3D_1+6\frac{C_1}{C_3}.
\end{equation}
This relation fixes the arbitrary parameter in the expression (\ref{solPsi1}) for the function $\Psi(t)$.
Thus, we get the function $f(\psi)$ which includes four arbitrary parameters $C_1$, $C_3$, $C_4$ and $D_1$. In particular, we have found that if $f(\psi)$ is an arbitrary quadratic polynomial, then the model has solutions with $H=1/(2t)$ if and only if $\Lambda=0$ and $\rho_{\mathrm{m}}(t)=0$.

Let us consider solutions with $H=1/(3t)$. Substituting the functions
$\psi(t)$, given by (\ref{1/3psi}), and  $\Psi(t)$ into
Eq.~(\ref{s}), we obtain the following solution:
\begin{equation}
\label{s1_3}
\begin{split}
s(t)&=\frac{1}{[2\ln(t)+3C_3]^2}\left(\frac {9}{4}[3{C_3}-1+2\ln(t)]\Lambda
t^2+{}\right.\\[2.7mm]
 &{}+\left.\frac{9\rho_0\kappa^2[(w_{\mathrm{m}}-1)(2\ln(t)
 +3C_3)+2]}{3w_{\mathrm{m}}-5}t^{1-w_{\mathrm{m}}}-\frac{2C_1[
3+2\ln(t)+3C_3]}{t^{2/3}}+\tilde{D}_1\right).
\end{split}
\end{equation}
Note that for $w_{\mathrm{m}}=1$ the term which is proportional to  $\rho_0$ can be actually considered as part of the integration constant $\tilde{D}_1$.

Equation~(\ref{einstein1}) can be rewritten in the following way
\begin{equation}
\label{einstein1s}
    3H^2\Psi=- \frac{1}{2}\left(\dot{\Psi}-s\dot{\psi}\right) \dot\psi
 - 3H\dot\Psi  + \Lambda
+ \kappa^2 \rho_{\mathrm{m}},
\end{equation}
therefore, the knowledge of the functions $H(t)$, $\Psi(t)$, $s(t)$, $\psi(t)$, and $\rho_{\mathrm{m}}(t)$ is enough in order to check the existence of solutions in the given form. One does not need to use the explicit form of $f(\psi)$. This property is useful in the case when the form  of $f(\psi)$ is too complicated, for example, when it is given by quadratures.
From (\ref{einstein1s}) we obtain the condition: \
$\tilde{D}_1=6C_2$.

After that, we use the relation between $t$ and $\psi$
\begin{equation}
t=e^{(-3\tilde{C}_3\pm J)/2},\qquad \mbox{where}\quad J=\sqrt{9\tilde{C}_3^2+12\psi-12\tilde{C}_4}\,,
\end{equation}
to get $f'(\psi)$. Note that the function $\psi(t)$ has a minimum, what means that we get two inverse functions.

 Integrating the function thus obtained over $\psi$, we get
\begin{eqnarray}
  \nonumber\qquad\qquad f(\psi)&=&-\frac{3\kappa^2\rho_0}
{3w_{\mathrm{m}}-5}e^{\frac32(w_{\mathrm{m}}-1)\tilde{C}_3}\left[e^{\mp\frac{(w_{\mathrm{m}}-1)J}{2}}
+\mathrm{Ei}\left(1,\pm\frac12(w_{\mathrm{m}}-1)J\right)\right]\\
&+&
C_1e^{\tilde{C}_3}\left[e^{\mp\frac{J}{3}}+\mathrm{Ei}\left(1,\pm\frac{J}{3}\right)\right]+{}\\&+&
\frac{3\Lambda
e^{-3\tilde{C}_3}}{8}\left[e^{\pm J}+\mathrm{Ei}\left(1,\mp
J\right)\right]+\frac{C_2}{2}\ln\left(3\tilde{C}_3^2+4\psi-4\tilde{C}_4\right)+\tilde{D}_2\nonumber,
\end{eqnarray}
where $\tilde{D}_2$ is a constant and the function $\mathrm{Ei}(1, -x)\equiv -\mathrm{Ei}(x)$,
where the exponential integral, for all non-zero values of $x$, is defined as follows:
\begin{equation*}
\mathrm{Ei}\left(x\right)= \int\limits_{-\infty}^x\frac{e^{z}}{z}\, d z.
\end{equation*}
For positive values of $x$ the integral has to be understood in terms of the Cauchy principal value, owing to the singularity of the integrand at zero. This example clearly shows that the method here considered allows to obtain the functions $f(\psi)$ even in the case when they are not elementary functions, what is a valuable bonus.  We also get the power-law solution in explicit~form.

\section{The reconstruction procedure for the model with a few perfect fluids}

In previous sections, we consider the model with one perfect fluid with constant parameter $w_m$. In this section, we show that the reconstruction procedure allows find models with power-law solutions, which include a few perfect fluids.
We assume that
\begin{equation*}
\rho_{\mathrm{m}}=\sum\limits_{i=1}^{N}\rho_{i},\qquad P_{\mathrm{m}}=\sum\limits_{i=1}^{N}P_{i},\qquad
P_{i}=w_i\rho_{i},
\end{equation*}
where $w_i$ are constants. For $H=n/t$, the corresponding continuity equations
\begin{equation}
\label{equ_rhoi}
\dot\rho_i={}-3H(w_i+1)\rho_i,
\end{equation}
have the following solutions:
\begin{equation}\label{rhoi}
    \rho_{i}(t)=\rho_{0i}t^{-3n(w_{i}+1)}.
\end{equation}

Substituting (\ref{rhoi}) into Eq.~(\ref{equPsi}), we get the following solution\footnote{We assume that values of $n$ and $w_i$ are such that we do not divide on zero.}
\begin{equation}
\label{solPsii}
   \Psi=C_1t^{-2n}+C_2t^{1-3n}+\frac{\Lambda}{(n+1)(3n+1)} t^2- \kappa^2\sum\limits_{i=1}^{N}\frac{\rho_{0i} (w_i-1)t^{2-3(1+w_i)n}}{(3nw_i-1)(n+3nw_i-2)}.
\end{equation}
After substitution the obtained function $\Psi(t)$ into Eq.~(\ref{equaf}) and using (\ref{tpsi}) we get
\begin{equation}
\label{fsumexp}
\begin{split}
 f_0(\psi)&=\frac{\Lambda t_0^2}{6n(1+n)}e^{(1-3n)\psi/(3n(2n-1))}\\&{}-\kappa^2\sum\limits_{i=1}^{N}\frac{\rho_{0i}t_0^{2-3n-3nw_{i}}}
 {3n(n-2+3nw_{i})}
e^{(3n(1+w_{i})-2)(3n-1)\psi/(6n(2n-1))}
\\&{}+\frac{(n-1)t_0^{-2n}C_1}{2(2n-1)}e^{(3n-1)\psi/(3(2n-1))}
+D_1e^{(3n-1)^2\psi/(3n(2n-1))}+D_2.
\end{split}
\end{equation}

To proof the existence of power-law solutions we substitute the functions obtained into Eq.~(\ref{einstein1}). After this substitution, we get that terms which are proportional to $\Lambda$ and $\rho_{0i}$ cancel and we get the condition
$D_1=0$, which coincide with the corresponding condition for the model with one perfect fluid (Eq.~(\ref{cond})).

Thus, we demonstrate that the reconstruction procedure can be used in the case of matter, which includes a few perfect fluids with constant EoS parameters, as well.

\section{Conclusion}

In this paper we have extended the reconstruction procedure for the scalar-tensor model proposed in~\cite{Odintsov0708}, which is a convenient local formulation of the popular nonlocal model~\cite{Deser:2007jk}.
We have explicitly shown that the models with $f(\psi)$ given by (\ref{fpsi}) with (\ref{fcoefs}) contain power-law solutions with $H=n/t$. The ones obtained include an arbitrary parameter, so that what we actually find is a full one-parameter family of solutions.
This result has been obtained for an arbitrary $n$, except for the values $n= -1$, $n= 1/3$, and $n= 1/2$. For those values of $n$ the function
$f(\psi)$ has a different form, which has been studied in detail, as well.

Another result of our work is that some models with $f(\psi)$ being a simple exponential function have both de Sitter and power-law solutions, but this occurs if and only if either
the cosmological constant, $\Lambda$, or the matter energy density, $\rho_{m}(t)$, are equal to zero. At the same time, the model with a single exponential $f(\psi)$, which does not include neither the cosmological constant nor matter contributions, can have either de Sitter solutions or power-law ones, but never both kinds simultaneously. These results characterize in a clear way all different possibilities. On top of these findings we have also demonstrated that there are models with both de Sitter and power-law solutions which include a cosmological constant and also matter terms. The function $f(\psi)$ for this kind of models is explicitly given by the sum of two exponential functions, specified  by~(\ref{fpsi2}) and (\ref{fcoefs2}).

In this paper, we consider the unperturbed universe, it is necessary to mention that only detailed study of perturbations in the modified gravity model is a potential way to get distinguish between the model and the general relativity. The perturbation analysis of the model without matter allow to get the constrains on the function $f(\psi)$ from the Solar System test~\cite{Koivisto2}, whereas a detailed study of matter perturbations and of their growth history is important to get constrains from the present and planning observations.  To the best of our knowledge the growth history index in nonlocal cosmological models has never been studied\footnote{After this paper has been published, we noticed the paper~\cite{Park} which deals with the evolution of the Universe structure
in the model described by the nonlocal action (\ref{action}).}. We plan to consider these problems for the models with the obtained functions $f(\psi)$ in future publications.

\medskip

\noindent {\bf Acknowledgements.}  The authors thank Sergei~D.~Odintsov
and Ying-li~Zhang for very useful discussions. E.E. was supported in
part by MICINN (Spain), grant PR2011-0128 and project FIS2010-15640, by
the CPAN Consolider Ingenio Project, and by AGAUR (Generalitat de
Ca\-ta\-lu\-nya), contract 2009SGR-994, and his research was partly carried out while on leave at the Department of Physics and Astronomy, Dartmouth College, NH, USA. The research of E.P. and S.V. is supported in
part by the RFBR grant 11-01-00894, E.P. by the RFBR grant 12-02-31109, and S.V. by the Russian Ministry of Education and Science under grant NSh-3920.2012.2, and by contract CPAN10-PD12 (ICE, Barcelona, Spain).

\end{document}